\documentclass[showpacs,twocolumn]{revtex4}

\usepackage{graphicx}
\usepackage{dcolumn}
\usepackage{amsmath}

\makeatletter
\def\btt#1{\texttt{\@backslashchar#1}}
\DeclareRobustCommand\bblash{\btt{\@backslashchar}} \makeatother
\begin{document}

\title{Strong lensing and Observables around 5D Myers-Perry black hole spacetime
}

\author{Ravi Shankar Kuniyal$^{a}$} \email{ravikuniyal09@gmail.com}
\author{Hemwati Nandan $^{a}$} \email{hnandan@associates.iucca.in}
\author{Uma Papnoi $^{b}$} \email{uma.papnoi@gmail.com}
\author{Rashmi Uniyal $^{c}$} \email{rashmiuniyal001@gmail.com}
\author{K D Purohit $^{d}$} \email{kdpurohit55@gmail.com}
\affiliation{$^{a}$ Department of Physics, Gurukula Kangri Vishwavidyalaya,\\
Haridwar 249 404, Uttarakhand, India}
\affiliation{$^{b}$ Kanoria PG Mahila Mahavidyalaya, Jaipur 302004, Rajasthan, India}
\affiliation{$^{c}$ Department of Physics, Govt. Degree College Narendranagar,\\ Tehri Garhwal - 249 175, Uttarakhand, India }
\affiliation{$^{d}$ Department of Physics, HNB Garhwal University,\\
Srinagar Garhwal 246 174, India}

\date{\today}

\begin{abstract}
We study the motion of massless test particles in a five dimensional (5D) Myers-Perry black hole spacetime with two spin parameters. The behaviour of the effective potential in view of different values of black hole parameters is discussed in the equatorial plane. The frequency shift of photons is calculated which is found to depend on the spin parameter of black hole and the observed redshift is discussed accordingly. The deflection angle and the strong deflection limit coefficients are also calculated and their behaviour with the spin parameters is analysed in detail. It is observed that the behaviour of both deflection angle and strong field coefficient differs from Kerr black hole spacetime in four dimensions (4D) in General Relativity (GR) which is mainly due to the presence of two spin parameters in higher dimension.
\keywords{5D Myers-Perry black hole; Null geodesics; Cone of avoidance; Frequency shift.}
\end{abstract}

\maketitle
\section{Introduction}
\label{intro}
The Black holes (BHs) in Einstein's General Relativity (GR) are one of the most strangest and mysterious 
objects in the universe \cite{har03,wal84,psj97,wei04}. The most general spherically symmetric, vacuum solution of the Einstein field equations in GR is the
well known Schwarzschild BH spacetime \cite{ep04,cha83} in four dimensions.  The study of Schwarzschild BH solution and its applications to the solar system is one of the accurate tests to verify the predictions made by GR. 
Further, a static solution to the
Einstein-Maxwell field equations, which corresponds to the gravitational field of a charged,
non-rotating, spherically symmetric body is the Reissner-Nordstr$\ddot{o}$m spacetime \cite{ep04,cha83}. The rotating generalization of the Schwarzschild black hole (BH) spacetime is Kerr BH spacetime in GR while the spacetime geometry in the region surrounding by a charged rotating BH is represented by the Kerr-Newman BH spacetime as a solution of Einstein-Maxwell equations in GR \cite{sch83,eva13}.\\
The GR which has revolutionized our understanding of the universe as a whole is now more than one hundred years old and the recent advancements in understanding the gravitational collapse and nature of BH solutions in diversified scenario is remarkable \cite{SMC,TP,GFR,AA}.\\
The deflection of light ray in a gravitational field is one of the crucial predictions of GR and the gravitational lensing is an important phenomena resulting due to the bending of light in the gravitational field of a massive object while passing close to that object. The strong gravitational lensing is caused by a compact objects like BHs with a photon sphere has distinctive features. It is worth mentioning that when the photons pass close to the photon sphere, the deflection angles become so
large that an observer would detect two infinite sets of faint relativistic images on each side of the BH,
which are produced by photons that make complete loops around the BH before reaching to the observer.
These relativistic images may therefore provide us not only some important signatures about BHs in the universe,
but might also helpful in verification of the alternative theories of gravity in their strong field regime. The gravitational lensing in weak field approximation studies the properties of galaxies and stars, but when BH is treated as a lens, it is no longer valid and the strong field limit is needed which is referred as strong deflection limit. Thus, it acts as a powerful indicator of the physical nature of the central celestial
objects and then has been used to study in various theories of gravity. The study of the strong field limit lensing due to different BHs have received considerable attentions in recent years \cite{sft,ksv,scht,vb,vb1,iye09,scht111}. The development of lensing theory in the strong-field regime started with the study of gravitational lensing due to a Schwarzschild BH spacetime \cite{sft,ksv} and it is also shown that a supermassive BH like at the center of our Galaxy may be a
suitable lens candidate \cite{ksv}.\\
In recent years, various interesting BH solutions in higher spacetime dimensions, especially in five dimensions \cite{upsgg}, have been the subject of intensive research, motivated by various ideas in brane-world cosmology, string theory and gauge/gravity duality \cite{horowitz}. It is worth to note that the uniqueness theorem does not hold in higher dimensions due to the fact that there are more degrees of freedom as compared to the usual four dimensions in GR. However, the discovery of
black-ring solutions in five dimensions asserts that the non-trivial
topologies are allowed in higher dimensions \cite{Emparan}. In particular, the
Myers-Perry black hole (MPBH) spacetime \cite{MP} is a higher dimensional
generalization of the four-dimensional Kerr BH spacetime in GR. 
The study of geodesic structure of massless particles in a given BH spacetime is one of the important ways to understand the gravitational field around a BH spacetime. 
The geodesic motion around various BH spacetimes in a variety of contexts (for timelike as well as null geodesics), both in GR and in alternating theories of gravity, are widely studied time and again \cite{dab97,fern12,fer12,fer14,nor05,pug11,hio08,rasprd,gus15,bha03,pra11,stu14,sch09,kol03,uni14,ras15,eva10,ru16,mak94,fer03,eva08,nan08,anv09,das09,das12,RefR11,ras17a,gho10,rav15,fuji09}. The motion of both massive and massless particles in Myers-Perry \cite{MP5Da} and Myers-Perry anti-de sitter BH spacetime \cite{MPADS} with equal rotation parameters
has been studied in detail. Further, the complete set of analytical solutions of the geodesic equations in the general 5D Myers-Perry spacetime in terms of the Weierstrass function, for the case of two independent angular momenta, have been derived and discussed in \cite{MP5Db}.  
Deimer et. al. also studied massive as well as massless test particles in the general 5D MPBH spacetime \cite{fuji09, MP5Da,MP5Db}. The main objective of this paper is to study the strong lensing in a 5D MPBH spacetime. We have calculated the deflection angle and other strong lensing parameters by using Bozza's method and the variation of deflection angle with spin parameter is investigated. We have used the units that fix the speed of light and the gravitational constant via $8\pi G = c^4 = 1$. \\ The paper is organised as follows: In Section II, the first integral of the geodesic equations and the effective potential in 5D MPBH spacetime are discussed. We have discussed the optical properties like frequency shift and the cone of avoidance from the null geodesics in Section III. The gravitational lensing aspects in the strong field limit is then discussed in detail in Section IV. Finally, the results obtained are summarised in Section V.
\section{Equations of Motion in 5D MPBH Spacetime}
\label{sec:1}
To study the geodesics and strong lensing in 5D rotating MPBH spacetime background, we begin with the following metric of MPBH spacetime in the Boyer-Lindquist coordinates \cite{upsgg},
\begin{flushleft}
\begin{eqnarray}\label{Kerr5D}
ds^2 &=&\frac{\rho^2}{ 4\Delta} dx^2+\rho^2
d\theta^2 - dt^2 +(x+a^2) \sin^2\theta  d\phi^2 \nonumber \\ && +  (x+b^2)  \cos^2\theta d\psi^2 \nonumber \\ && + \frac{2m}{\rho^2} \Big[dt+a \sin^2\theta d\phi  + b
\cos^2\theta d\psi  \Big]^2, 
\end{eqnarray}
\end{flushleft}
with $\rho^2$ and $\Delta$ are defined as
\begin{eqnarray}
\rho^2=x+a^2\cos^2\theta+b^2 \sin^2\theta ,\\
\Delta=(x+a^2)(x+b^2)-2\hspace{0.1cm}m x. 
\end{eqnarray}
The metric (\ref{Kerr5D}) is singular when $\Delta=g_{rr}=0$ and $\rho^2=0$. Here $a$ and $b$ are two spin parameters, and $\phi$ and $\psi$ are angles bounded by the limit $0\leq\phi\leq2\pi$ and $0\leq\psi\leq\pi/2$. Following \cite{Fr03}, we use the coordinate $x=r^2$ instead of the radius $r$ in order to simplify the calculations. It is worth noticing here that the metric (\ref{Kerr5D}) reduces to 5D Tangherlini solution \cite{scht} for $a=b=0$.
For the horizon structure of MPBH sapcetime with $\Delta=0$, we obtain
\begin{equation}\label{horizon}
x_{\pm}=\frac{1}{2}\left[2 m-(a^2 + b^2)\pm
\sqrt{[2 m-(a^2 + b^2)]^2 - 4 a^2 \hspace{1mm} b^2}\right] .\nonumber
\end{equation}Here, $x_+$ and $x_-$ denotes the outer horizon and the inner horizon respectively. The metric (\ref{Kerr5D}) describes non-extremal BH for $x_+>x_-$ and when $x_+=x_-$, one can obtain an extremal BH spacetime.  The horizon exists when $a^2+b^2<2m$ and $[2m-(a^2+b^2)]^2\geq4 a^2b^2$. \\
For the metric (\ref{Kerr5D})\begin{equation}
\sqrt{-g}=\frac{1}{2}\sin\theta \cos\theta \rho^2. \nonumber
\end{equation}
To study the geodesic structure of the 5D
rotating Myers-Perry black hole, we begin with the Lagrangian
which reads \begin{equation} L = \frac{1}{2} g_{\mu
\nu}\dot{x}^{\mu}\dot{x}^{\nu},
\end{equation}where an overdot denotes the partial derivative with respect to an affine parameter. Therefore, the momenta calculated for the metric (\ref{Kerr5D}) are:
\begin{eqnarray}
p_t &=& \left(-1+\frac{r_0^2}{\rho^2}\right)\dot{t}+\frac{r_0^2}{\rho^2}\dot{\phi}+\frac{r_0^2 b \cos^2\theta}{\rho^2}\dot{\psi}, \nonumber \\
p_{\phi}&=&\frac{r_0^2 a \sin^2\theta}{\rho^2}\dot{t}+\left(x+a^2+\frac{r_0^2 a^2 \sin^2\theta}{\rho^2}\right)\sin^2\theta \dot{\phi} \nonumber \\ && + \frac{r_0^2 a b \sin^2\theta \cos^2\theta}{\rho^2} \dot{\psi}, \nonumber \\
p_{\psi}&=& \frac{r_0^2 b \cos^2\theta}{\rho^2}\dot{t} + \frac{r_0^2 a b \sin^2\theta \cos^2\theta}{\rho^2} \dot{\phi} \nonumber \\ &&  + \left(x+b^2+\frac{r_0^2 b^2 \cos^2\theta}{\rho^2}\right)\cos^2\theta \dot{\psi}, \nonumber \\
p_x &=& \frac{\rho^2}{4\Delta}\dot{x}, \nonumber \\
p_\theta &=& \rho^2 \dot{\theta},
\end{eqnarray}where $p_t = -E$, $p_{\phi} = L_{\phi}$ and $p_{\psi}=L_{\psi}$ correspond to  energy and angular momentum with respect to the respective rotation axis  respectively.
\\
Now considering the case for the equatorial plane, i.e., $\theta=\pi/2$, which results in the conserved quantity along $\psi$ direction, i.e., $L_{\psi}=0$. Henceforth, we will be using $L_{\phi} = L$ in our calculation. Thus the metric (\ref{Kerr5D}) in the equatorial plane reads as
\begin{equation}\label{spacetime1}
ds^2 = - A(x) \hspace{0.1cm}dt^2 + B (x)\hspace{0.1cm} dx^2 + C(x) \hspace{0.1cm}d\phi^2 - D (x) \hspace{0.1cm}dt\hspace{0.1cm} d\phi,
\end{equation}
where the metric coefficients are described as below,
\begin{eqnarray}
A(x) &=& (1-\frac{2 m}{\rho^2}),\\ 
B(x) &=& \frac{\rho^2}{4 \Delta},\\ 
C(x) &=& x + a^2 + \frac{2 m a^2}{\rho^2},\\
D(x) &=& \frac{-4 m a}{\rho^2}.
\end{eqnarray}
The first integral of geodesic equations may then be expressed in terms of the above mentioned metric coefficients \cite{vb}, in the following form,
\begin{eqnarray}
\dot{t} &=& \frac{4 C(x) E - 2 D(x) L}{4 A(x) C(x) + D(x)^2}, \label{EOM1}\\
\dot{x} &=& \pm 2 \sqrt{\frac{C(x) E^2 - D(x) E\hspace{1mm}L - A(x) L^2}{B (4 \hspace{1mm}A(x)\hspace{1mm} C(x) + D(x)^2)}}, \label{EOM2} \\
\dot{\phi} &=& \frac{2 D(x) E + 4 \hspace{1mm}A(x)\hspace{1mm}L}{4\hspace{1mm}A(x)\hspace{1mm}C(x) + D(x)^2}. \label{EOM3}
\end{eqnarray}
For null geodesics, $\dot{x}$ from Eq. (\ref{EOM2}), can be reconstructed as
\begin{equation}
\dot{x}^2 + V_{eff} = 0,
\end{equation}
which gives,
\begin{eqnarray}
V_{eff} &=& - 4 \Big[\frac{C(x)E^2 - D(x)EL - A(x) L^2}{B(x) (4 A(x)C(x) + D(x)^2)}\Big], \nonumber \\
        &=& \frac{1}{x+b^2}\Big( -4 E^2(2 m a^2 + x^2 + x b^2 + a^2 x + a^2 b^2) 
       \nonumber \\ && - 16 m a E L + 4 L^2(x+ b^2 - 2 m)\Big).\nonumber \\
\end{eqnarray}
\begin{figure}
\resizebox{0.35\textwidth}{!}{%
\includegraphics{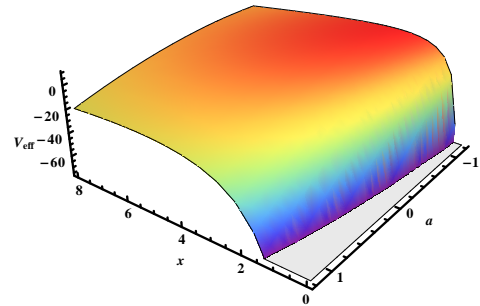}
}
\hspace{3cm}
\resizebox{0.35\textwidth}{!}{%
\includegraphics{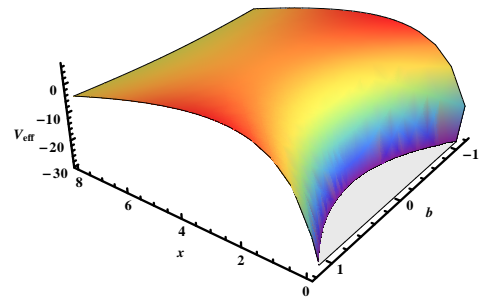}
}
\caption{\label{Veff} Variation of effective potential with radius (a) at different values of spin parameter $b$ for a fixed value of spin parameter $a (= 0.1)$ (b)  at different values of spin parameter $a$ for a fixed value of spin parameter $b(= 0.1)$. Here, $E = 1$ and $L = 3$.}
\end{figure}
The general behavior of effective potential as a function of $x$ for different values of rotation parameter is presented in Fig. (\ref{Veff}). In particular, Fig. (\ref{Veff}a) represents the variation of potential with the spin parameter $b$ for  fixed value of $a (=0.1)$ while Fig. (\ref{Veff}b) represents the variation of the potential with spin parameter $a$ for  fixed value of $b (=0.1)$. The effective potential shows a maxima which corresponds to an unstable circular orbit. It is also observed that with the increase in the value of parameter $b$, the maximum of the effective potential is shifting towards the left (see Fig. (\ref{Veff}a)), i.e., the circular orbits also shift towards the central object accordingly whereas with the increase in the value of spin parameters $a$ at fixed $b$, the peak is shifting towards the right (see Fig.(\ref{Veff}b)), which signifies the shifting of circular orbit away from the central object.
\section{Observables for Photons}
\label{sec:2}
In order to discuss the optical properties in 5D MPBH spacetime, the frequency shift and cone of avoidance are discussed below:
\subsection{Frequency shift}
The angular frequency associated with photons in a circular geodesic is one of the meaningful physical quantity. The angular frequency relative to a distant observer for unequal spin parameters is defined as below,
\begin{equation}
\Omega=\frac{d\phi}{dt}. \label{fs}
\end{equation}
Thus, using Eq. (\ref{EOM1}) and Eq. (\ref{EOM3}), the angular frequency given by Eq. (\ref{fs}) can be calculated as below,
\begin{equation}\label{angfre}
\Omega =\frac{(x+b^2-2 m) d - 2 m a}{\left( x^2 + (a^2 + b^2 - 2 m) x + a^2 b^2 + 2 m a \right) d + 2 m x + 2 m a^2 }, 
\end{equation}
where $d=L/E$. The frequency shift may however be expressed as 
\begin{equation}\label{frsh}
g = \frac{k_{\mu}u_o^{\mu}}{k_{\mu}u_e^{\mu}},
\end{equation}where $k_{\mu}$ are the covariant components of the photon four-momentum and $u_o^{\mu}$ ($u_e^{\mu}$, respectively) are the contravariant components of the four-velocity of the observer (emitter). In case of static distant observer, the four-velocity reads $u_o = (1,0,0,0,0)$ and in the case of emitter the four-velocity reads as $u_e = (u_e^{t},0,0,u_e^{\phi},0)$. The frequency shift then acquires the following form,
\begin{equation}
g= \frac{1}{u_e^{t}(1- d\Omega)}.
\end{equation}The temporal component of the emitter four-velocity can be obtained from the norm of the four-velocity:
\begin{equation}
u_e^{t} = \left[1-\frac{2m}{\rho^2}-\left(x+a^2+\frac{2ma^2}{\rho^2}\right)\Omega^2\right]^{-1/2},
\end{equation}
such that the expression for frequency shift now reads as,
\begin{equation}\label{FS}
g = \frac{\left[1-\frac{2m}{\rho^2}-\left(x+a^2+\frac{2ma^2}{\rho^2}\right)\Omega^2\right]^{1/2}}{1-d\Omega}.
\end{equation} 
Here, by considering the value of $a$ in between $0$ and $1$, one automatically has a range for $b$ (from the expressions $a^2+b^2 <2m$ and $[2m-(a^2+b^2)]^2 \geq 4a^2b^2$) with $m=1$ as, $-0.4 \geq b \geq 0.4$ or $-2.412 \geq b \geq 2.412$. 
\begin{figure}
\resizebox{0.35\textwidth}{!}{%
\includegraphics{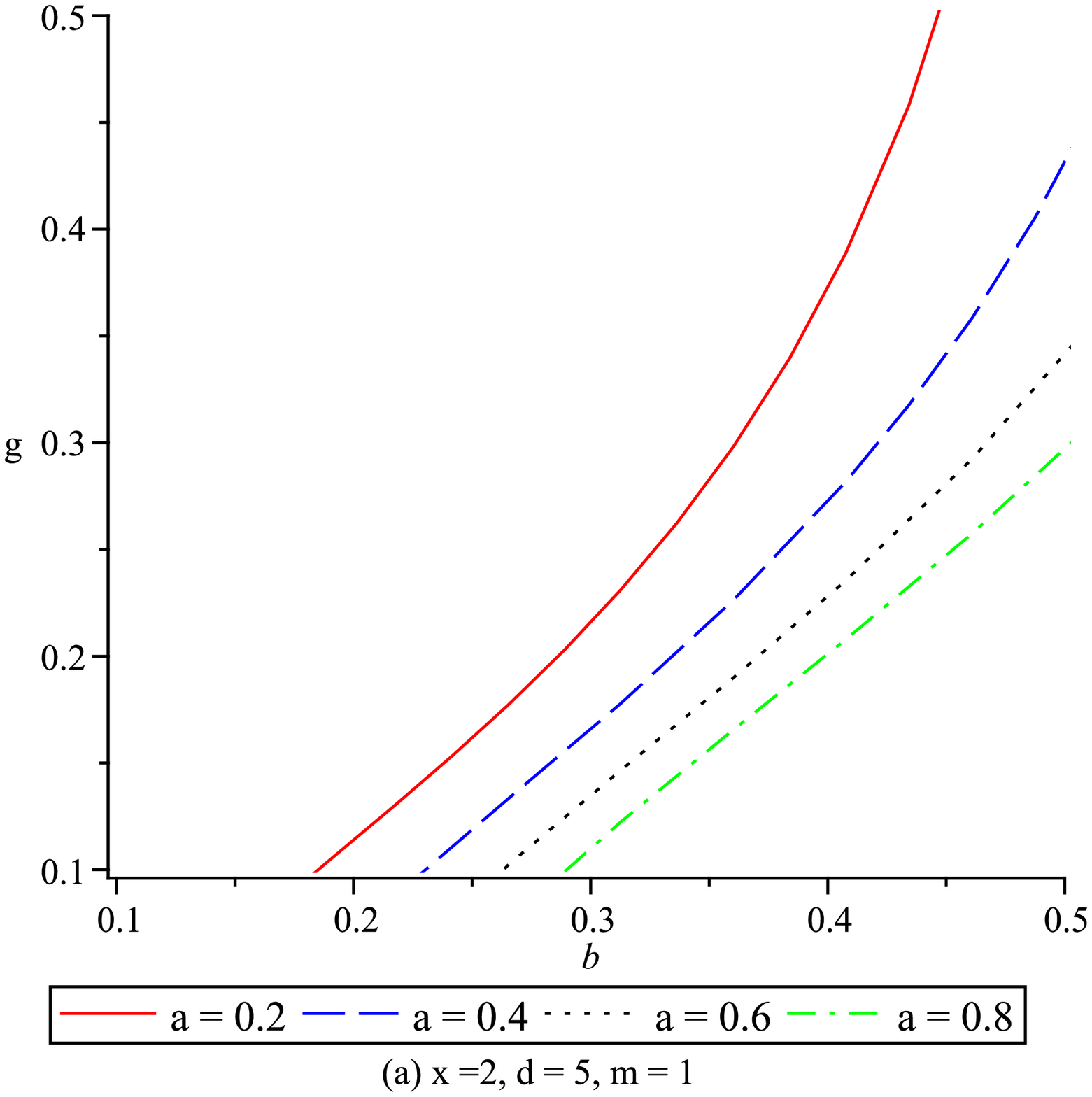}}
\hspace{3cm}
\resizebox{0.35\textwidth}{!}{%
\includegraphics{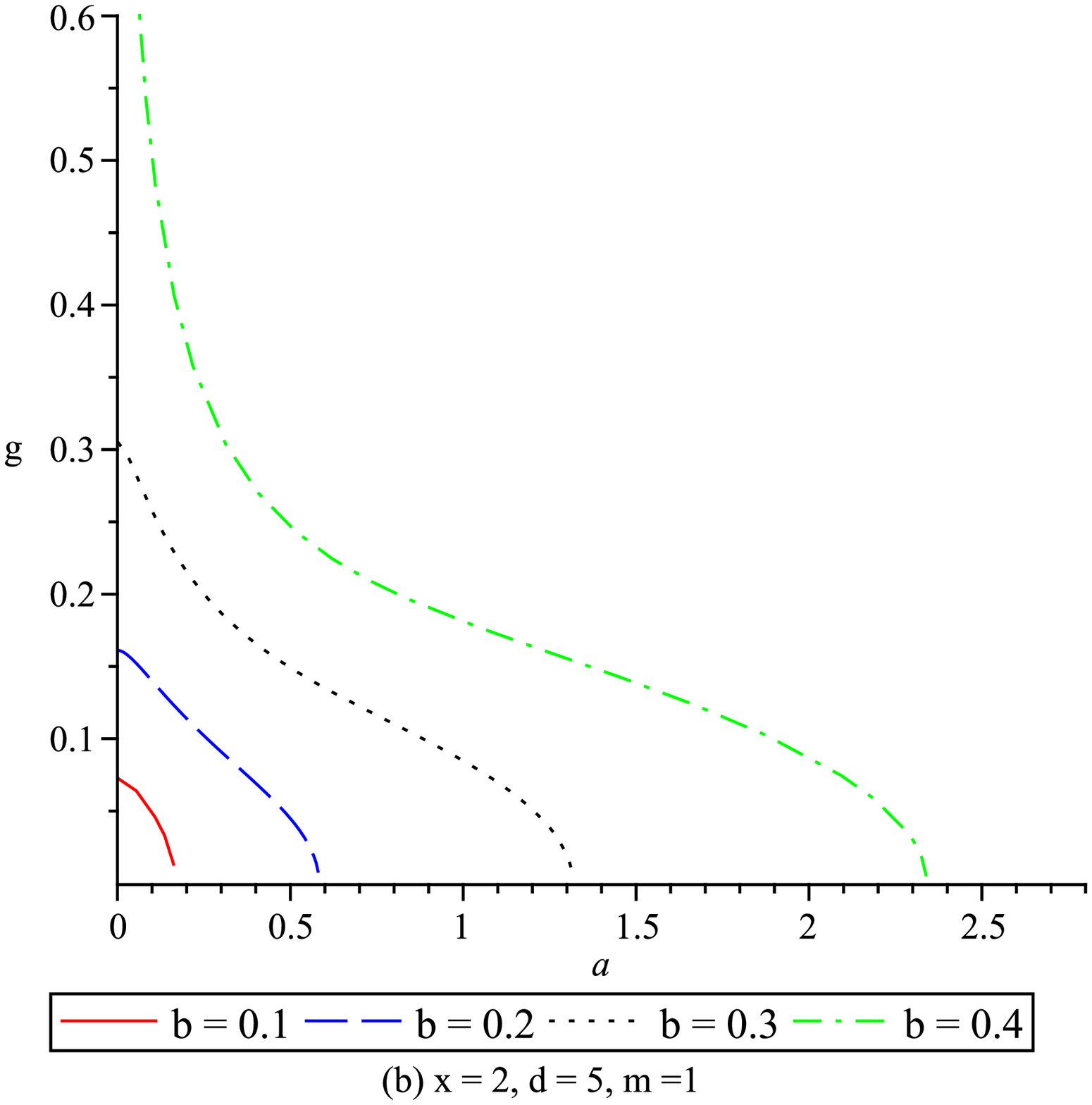}}
\caption{\label{fs1} Variation of frequency shift: (a) with spin parameter $b$ for different values of spin parameter $a$, (b)  Variation of frequency shift with spin parameter $a$ for different values of spin parameter $b$.}
\end{figure}
\begin{figure}
\resizebox{0.35\textwidth}{!}{%
\includegraphics{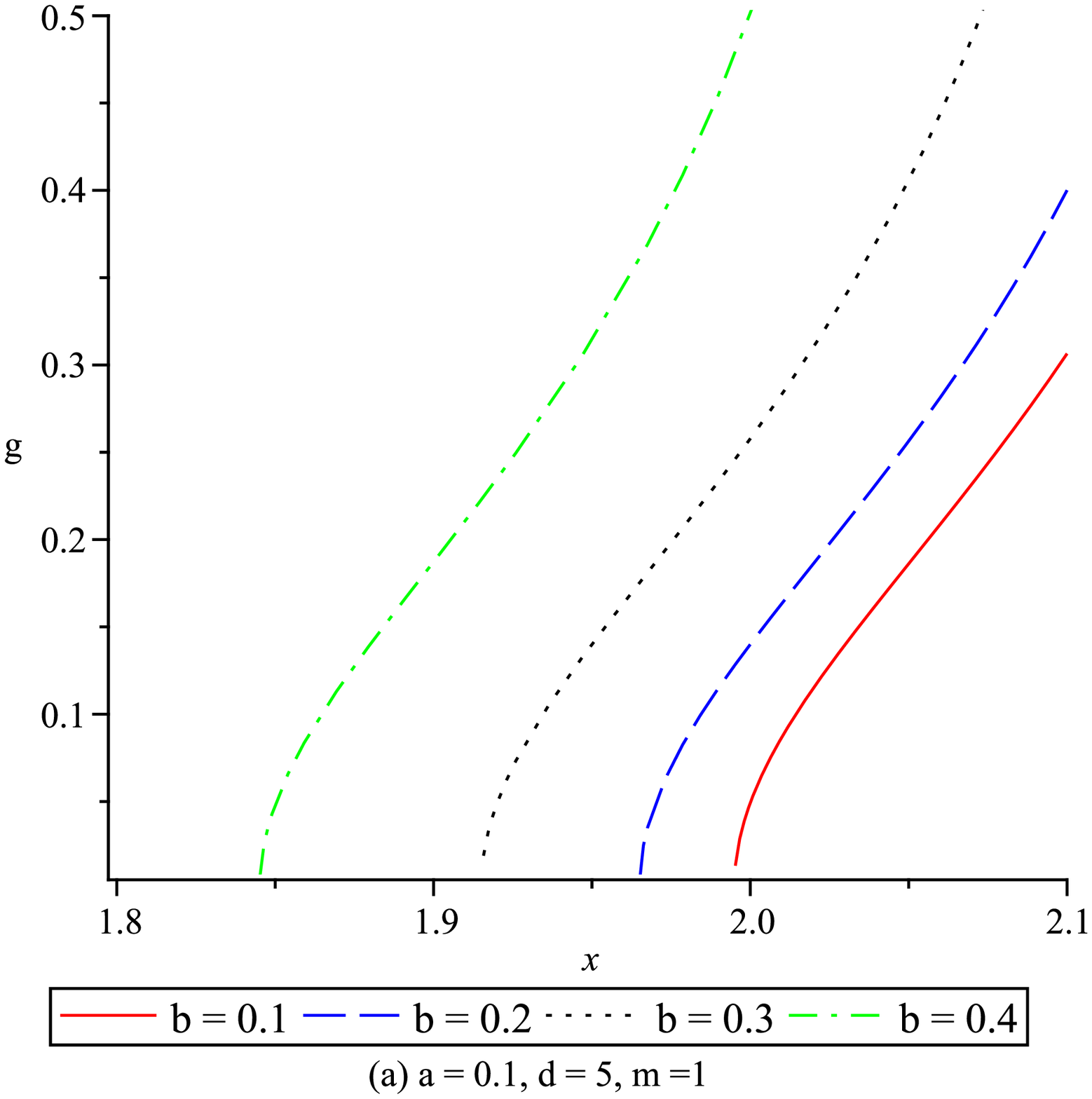}}
\hspace{3cm}
\resizebox{0.35\textwidth}{!}{%
\includegraphics{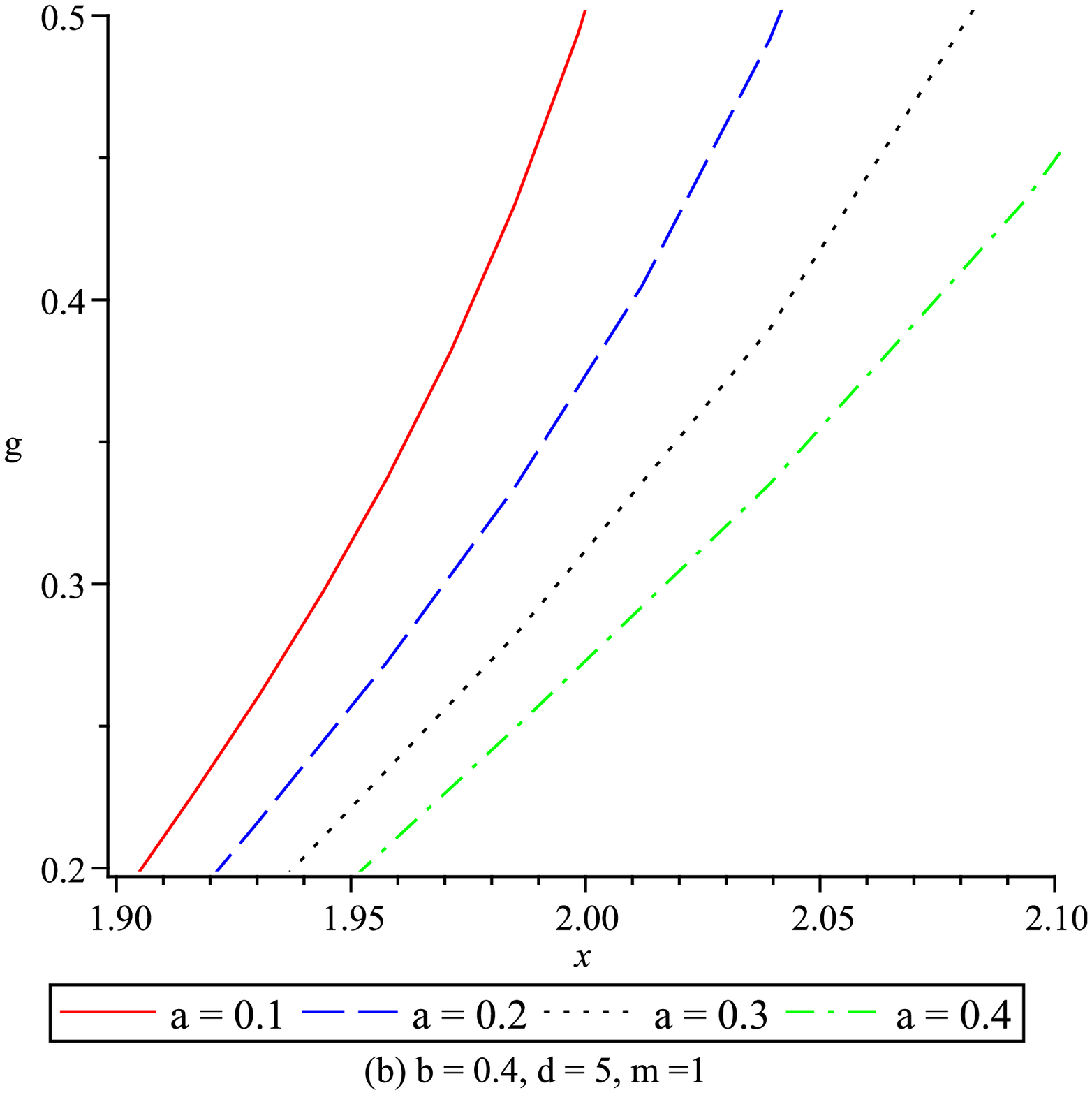}}
\caption{\label{fs2} Variation of frequency shift with $x$: (a) for different values of spin parameter $b$, with $a=0.1$ and (b) for different
values of spin parameter $a$, with $b=0.4$.}
\end{figure}
\begin{figure}
\begin{center}
\resizebox{0.35\textwidth}{!}{%
\includegraphics{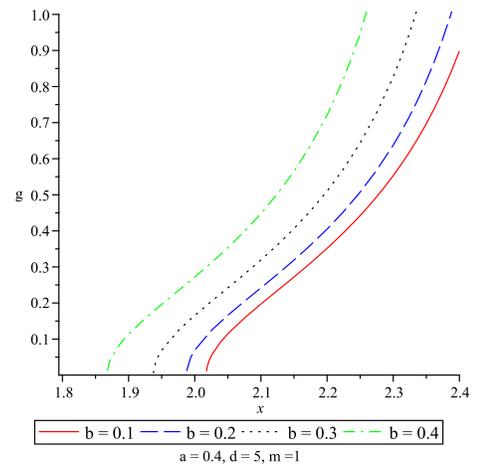}}
\caption{\label{fs3} Variation of frequency shift with $x$ for different values of $b$, with $a=0.4$. }
\end{center}
\end{figure}
The behaviour of the frequency shift for different values of different spin parameters is illustrated in Figs. (\ref{fs1})- (\ref{fs3}). In particular in Fig. (\ref{fs1}), we have presented the frequency shift for different values of spin parameter $a$ and $b$ while keeping the other parameter constant respectively. The Fig. (\ref{fs1}a), shows the variation of frequency shift with spin parameter $b$ at different values of spin parameter $a$, whereas Fig. (\ref{fs1}b), shows the variation of frequency shift with $a$ at different values of $b$. As the frequency shift increases, the observed frequency of the photon decreases which in turn gives an equivalent increase in the corresponding wavelength of the photon. Hence, for the increasing frequency shifts, the photons get redshifted. One may note that the stronger is the gravitational field of a source, larger will be the energy loss of an incoming photon and also larger will be the observed redshift.\\
From Fig. (\ref{fs1}), it can also be observed that for a particular value of parameter $a$, the redshift for photons around a MPBH spacetime increases with an increase in the value of spin parameter $b$ while it decreases with an increase in the value of parameter $a$. It therefore signifies the strength of gravitational field which depends strongly on both the
parameters at a fixed value of $x$. In Fig. (\ref{fs2}) and  Fig. (\ref{fs3}), the variation of frequency shift with $x$ for different values of spin parameters $a$ and $b$ is shown respectively. It is also observed  that for a particular value of spin parameter $b$, there is a decrease in the redshift for the photons with an increase in the value of $x$. However, for fixed value of $a$, the redshift first increases and then decreases with $x$.
\subsection{Null geodesics in the observers frame}
\noindent Let us consider a point source at a given distance $r$ from the centre emits light isotropically into all directions.  A part of the light will then be captured by the BH, while another part will escape from the vicinity of BH. It is clear that the critical orbits are at the limit of infall or escape and therefore such orbits may define a cone with half opening angle  $\psi$ in the observer's frame \cite{gus15} such that,
\begin{equation}
\tan \psi = \frac{k_{x}}{k_{\phi}}.
\label{coa}
\end{equation}
Here, $k_x$ and $k_{\phi}$ are the components of the null vector field $k = k^{\mu}\partial_{\mu}$ for null geodesic in this spacetime. In general case, the null vector field $k$ has four components ($k^t$, $k^x$, $k^{\theta}$, $k^\phi$). Using the equations for constant of motion, i.e., $E = -g_{t\mu} k^{\mu}$ and $L = g_{\phi \mu} k^{\mu}$, we obtain the relations,
\begin{equation}
k^t = \frac{G_t(E,L)}{G}, \hspace{2cm} k^{\phi} = \frac{G_{\phi}(E,L)}{G},
\end{equation}
where, $G =  A(x) C(x) + \frac{ D(x)^2}{4}$ and the functions $G_t$ and $G_\phi$ are defined as below,
\begin{eqnarray}
G_t(E,L) =  C(x) E - \frac{ D(x) L}{2},\\
G_{\phi}(E,L) = \frac{D(x) E}{2}  +  A(x) L.
\end{eqnarray}
Using the constraint $k\cdot k = 0$ (i.e $(k^t)^2 + (k^x)^2 + (k^{\theta})^2 + (k^{\phi})^2 = 0$) for null geodesics, we can then obtain an expression for $k^x$ in equatorial plane,
\begin{equation}
(k^x)^2  = \frac{W}{B(x) G},
\end{equation}
where $W = C(x) E^2 - D(x) E L - A(x) L^2$.\\
Further, $k_{\alpha} = g_{\mu \nu} k^{\mu} u^{\nu}_{(\alpha)}$ leads to the components of $k$ in observers frame as follows,
\begin{eqnarray}
k_{x} &=& \frac{G_t}{\sqrt{- G C(x)}} \sinh q 
        -  
  \left( \sqrt{\frac{-W}{G}} \cos \chi - \frac{L}{\sqrt{-C(x)}} \sin \chi \right)\nonumber \\ && \cosh q, \label{kx}
\end{eqnarray}
\begin{equation}
k_{\theta} = 0, \hspace{0.5cm} k_{\phi} = \sqrt{\frac{-W}{ G }} \sin \chi + \frac{L}{\sqrt{-C(x)}} \cos \chi. \label{kphi}
\end{equation}
Using Eqs. (\ref{kx}) and (\ref{kphi}) in Eq.(\ref{coa}), one can then obtain
\begin{widetext}

\begin{equation}
\tan \psi = \frac{\frac{G_t}{\sqrt{G}} \sinh q - \left(\sqrt{\frac{ W \hspace{0.1cm} C(x)}{G}} \cos \chi - L \sin \chi\right)\cosh q}{\left(\sqrt{\frac{ W \hspace{0.1cm} C(x)}{G}} \sin \chi + L \cos \chi\right)}.
\label{coa_1}
\end{equation}

\end{widetext}
Here, we have considered the null geodesics in observer's frame moving with constant acceleration. However an arbitrary observer in the spacetime (\ref{spacetime1}) is determined by a velocity vector field $u$, (i.e., $u^2 = -1$ or $+1 $, respectively for timelike and null geodesics) as,
\begin{equation}
u = \alpha \partial_t  + \beta \partial_x + \gamma \partial_{\phi}.
\label{vector field_1}
\end{equation}
The vector field can be parametrized by a pair of function $(q(x), \chi(x))$ (see \cite{gus15}) as given below,
\begin{eqnarray}
u &=& \sqrt{\frac{-C(x)}{G}} \cosh q  \partial_t + \frac{1}{\sqrt{-B(x)}} \sinh q \cos \chi \partial_x \nonumber \\ && + \frac{1}{\sqrt{-C(x)}} \Big(\sinh q \sin \chi - \frac{D(x)}{2 \sqrt{G}}  \cosh q \Big)\partial_{\phi}.\nonumber \\
\label{vector field_2}
\end{eqnarray}
The other basis vectors $u_i$ are,
\begin{eqnarray}
u_x &=& \sqrt{\frac{-C(x)}{G}} \sinh q  \partial_t + \frac{1}{\sqrt{-B(x)}} \cosh q \cos \chi \partial_x  \nonumber \\ && + \frac{1}{\sqrt{-C(x)}} \left(\cosh q \sin \chi - \frac{D(x)}{2 \sqrt{G}}  \sinh q  \right)\partial_{\phi},\label{vector field_4} \nonumber \\
u_{\phi} &=& - \frac{\sin \chi}{\sqrt{-B(x)}} \partial_x + \frac{\cos \chi}{\sqrt{-C(x)}} \partial_{\phi}.
\label{vector field_4}
\end{eqnarray}
The (equatorial) trajectories of the observer (\ref{vector field_2}) are given by following equations,
\begin{eqnarray}
\frac{dt}{d\lambda} &=& \sqrt{\frac{-C(x)}{G}} \cosh q,\label{integral eqn1}\\
\frac{dx}{d\lambda} &=& \sqrt{\frac{1}{-B(x)}} \sinh q \cos \chi,\label{integral eqn2}\\
\frac{d\theta}{d\lambda} &=& 0,\label{integral eqn3}\\
\frac{d\phi}{d\lambda} &=& \frac{1}{\sqrt{-C(x)}} \left(\sinh q \sin \chi - \frac{D(x)}{2 \sqrt{G}} \cosh q \right),
\label{integral eqn4}
\end{eqnarray}
where $\lambda$ is an affine parameter.
From Eqs. (\ref{integral eqn1})-(\ref{integral eqn4}), the static observer (for  $\chi = \pi/2$ and $\tanh q = \frac{D(x)}{2 \sqrt{G}}$) is determined as follows,
\begin{eqnarray} 
\cosh q = \frac{2 \sqrt{G}}{\sqrt{4 G - D(x)^2}}, \\
 \sinh q = \frac{D(x)}{\sqrt{4 G - D(x)^2}}. 
\end{eqnarray}
The radial trajectories ($\theta =$ constant) are given by the conditions,
\begin{equation}
\tanh q \sin \chi = \frac{D(x)}{2 \sqrt{G}}, \hspace{2cm} \chi \neq 0.
\end{equation}
For the radial trajectories of the observer in equatorial plane i.e., $\chi = \pi/2$, $\tanh q  = \frac{D(x)}{2 \sqrt{G}}$. However for $\chi = \pi/4$, the relation $ \sinh q = \frac{D(x)}{\sqrt{2 G}} \cosh q$ holds and therefore the observer's velocity vector $u$ takes the following form,
\begin{equation}
u = \sqrt{\frac{-C(x)}{G}} \cosh q \hspace{1mm} \partial_t + \frac{D(x)}{2 \sqrt{-B(x) G}} \cosh q \hspace{1mm} \partial_x ,
\end{equation}
and from the condition of velocity vector field one can easily obtain $\cosh q = \frac{ \sqrt{G}}{\sqrt{ A(x) C(x) - \frac{D(x)^2}{4} }}$.
Thus the expression (\ref{coa_1}) now reads as,
\begin{widetext}
\begin{eqnarray}
\tan \psi &=& \frac{1}{\sqrt{G (  A(x) C(x) - \frac{D^2(x)}{4} )}} \times  \frac{G_t D(x) - \left( \sqrt{\frac{W C(x)}{G}} -  L\right) G}{\left( \sqrt{\frac{W C(x)}{G}} +  L\right)},
\end{eqnarray}
\end{widetext}
which clearly indicates that the angle $\psi$ depends on the two parameters of the null geodesics, i.e., $E$ and $L$.
\section{Strong field lensing by 5D MPBH Spacetime}
In this section, we investigate the strong field lensing by a 5D MPBH spacetime given by Eq. (\ref{Kerr5D}) for the case where both the observer and the source lie in the equatorial plane i.e., $\theta = \pi/2$ \cite{vb,vb1}. The impact parameter is related to the minimum distance reached by the photon. 
In general, a light ray coming from infinity approaches the BH, reaches the minimum
distance and then leaves again towards infinity. Using the geodesic equations, we find an implicit relation between angular momentum and the closest approach distance. 
Here for simplicity, we are considering $E = 1$, and at the minimum distance (say $x_0$) of photon trajectory (where $V_{eff} = 0$), we have \cite{vb},
\begin{eqnarray}
L &=& \frac{- D(x_0) + \sqrt{4 A(x_0) C(x_0) + D(x_0)^{2}(x)}}{2 A(x_0)}, \nonumber \\
  &=& {\frac {2\,m a - \rho\,\sqrt {x_0{\rho}^{2}+{a}^{2}{\rho}^{2}-2\,m x_0}}{2\,m - {
\rho}^{2}}}.
\end{eqnarray}
Now following \cite{vb1}, the condition for the radius of photon sphere is,
\begin{eqnarray}
&& A(x_p) C^{'}(x_p) - A^{'}(x_p) C(x_p) +  L(x_p) \Big(A^{'}(x_p)D(x_p)\nonumber \\ && - A(x_p) D^{'}(x_p)\Big) = 0. \nonumber \\
\end{eqnarray}
Thus, the equation for the radius of photon sphere takes the following form,
\begin{eqnarray}
&&{x}_{p}^{4}+ \left( -8\,m+4\,{b}^{2} \right) {x}_{p}^{3} + \left( 6\,{b}^{4}-20
\,m{b}^{2}+16\,{m}^{2}-8\,m{a}^{2} \right) \nonumber \\ 
&& {x}_{p}^{2} - \left( 16\,m{b}^{
4}- 4\,{b}^{6} - 16\,{b}^{2}{m}^{2} + 16\,{b}^{2}m{a}^{2} \right) x_{p} + {b}^{8}\nonumber \\ && -8\,{b}^{4}m{a}^{2} -4\,{b}^{6}m +4\,{b}^{4}{m}^{2}=0. \nonumber \\
\label{eqn:photon sphere}
\end{eqnarray}In the limit $a=0$ and $b=0$, the radius of the photon sphere comes out as $x_p = 4 m$, i.e., the radius of photon sphere for the Tangherlini spacetime \cite{scht111}.
\begin{figure}
\begin{center}
\resizebox{0.35\textwidth}{!}{%
\includegraphics{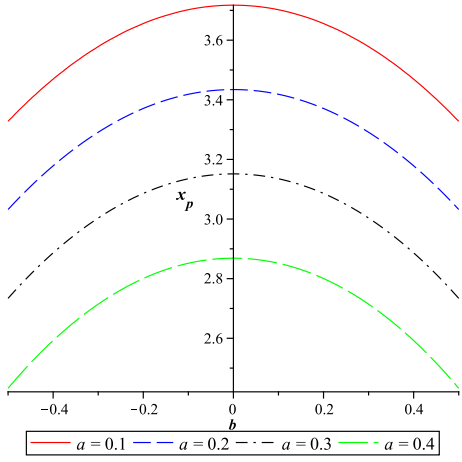}}
 \caption{\label{f9} Variation of radius of photon sphere as a function of spin parameter $b$ for different values of spin parameter $a$ with $m = 1$.}
 \end{center}
\end{figure}
\noindent The radius of the photon sphere is plotted with respect to the spin parameter $b$ in Fig. (\ref{f9}) for different values of spin parameter $a$. One can easily notice from the plots that on increasing the value of spin parameter $a$ the radius of photon sphere decreases.\\
Depending on the direction of the rotation of BH there are two types of photon spheres, one for the photons winding in the same direction of rotation as the BH known as direct photons and the other one for photons winding in the opposite direction known as retrograde photons. One may also notice that both direct and retrograde photons have same impact, i.e., in both the cases the photons are not easily captured by increasing the value of spin parameter $a$.\\
The deflection angle for photons coming from infinity can be written as,
\begin{equation}\label{aa}
\alpha(x_0) = \phi(x_0) - \pi,
\end{equation}
where $\phi(x_0)$ is the total azimuthal angle, which evaluates to $\pi$ for a straight line and becomes larger  with the bending of light ray in gravitational field. As the distance of closest approach $x_0$ decreases, the deflection angle increases accordingly. When $x_0$ reaches a minimum value, i.e.,  the radius of the photon sphere, the deflection angle becomes very large, and a photon will be captured by the BH.
\begin{figure}
\begin{center}
\resizebox{0.35\textwidth}{!}{%
\includegraphics{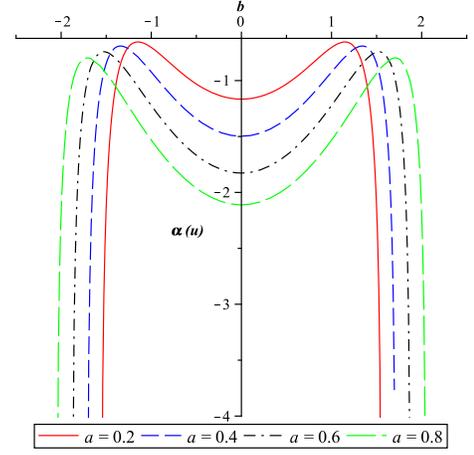}}
 \caption{\label{f12} Variation of deflection angle with spin parameter $b$ for different value of spin parameter $a$ with $m = 1$.}
 \end{center}
\end{figure}
 Now, using Eq. (\ref{aa}), azimuthal angle is given by,
\begin{eqnarray}
\phi(x_0)=2 \int_{x_{0}}^{\infty} \frac{d\phi}{dx}dx, \nonumber \\ \end{eqnarray}
\begin{eqnarray}
=2 \int_{x_{0}}^{\infty} \frac{\sqrt{B(x) \vert A(x_0) \vert} (D(x) + 2 L A(x))}{\sqrt{4 A(x) C(x) + D^2(x)} \sqrt{sgn(A(x_0))\hspace{1mm} P}} dx, \nonumber
\end{eqnarray}
where,
\begin{eqnarray}
P &=&  C(x) A(x_0) - A(x) C(x_0) 
        + L \Big[A(x) D(x_0) \nonumber \\ &&- A(x_0) D(x)\Big].
\end{eqnarray}
\begin{figure}
\resizebox{0.35\textwidth}{!}{%
\includegraphics{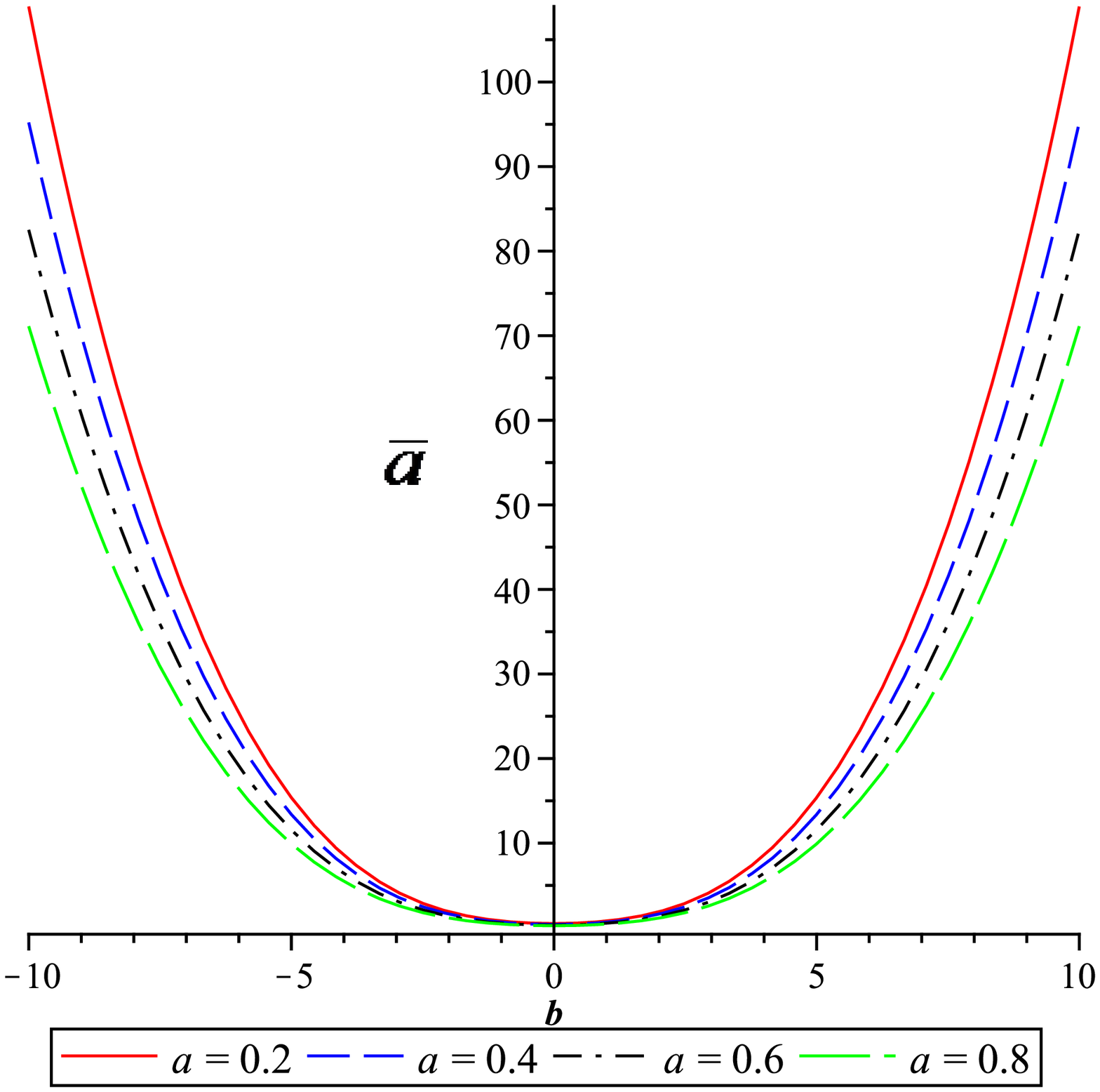}}
\hspace{3cm}
\resizebox{0.35\textwidth}{!}{%
\includegraphics{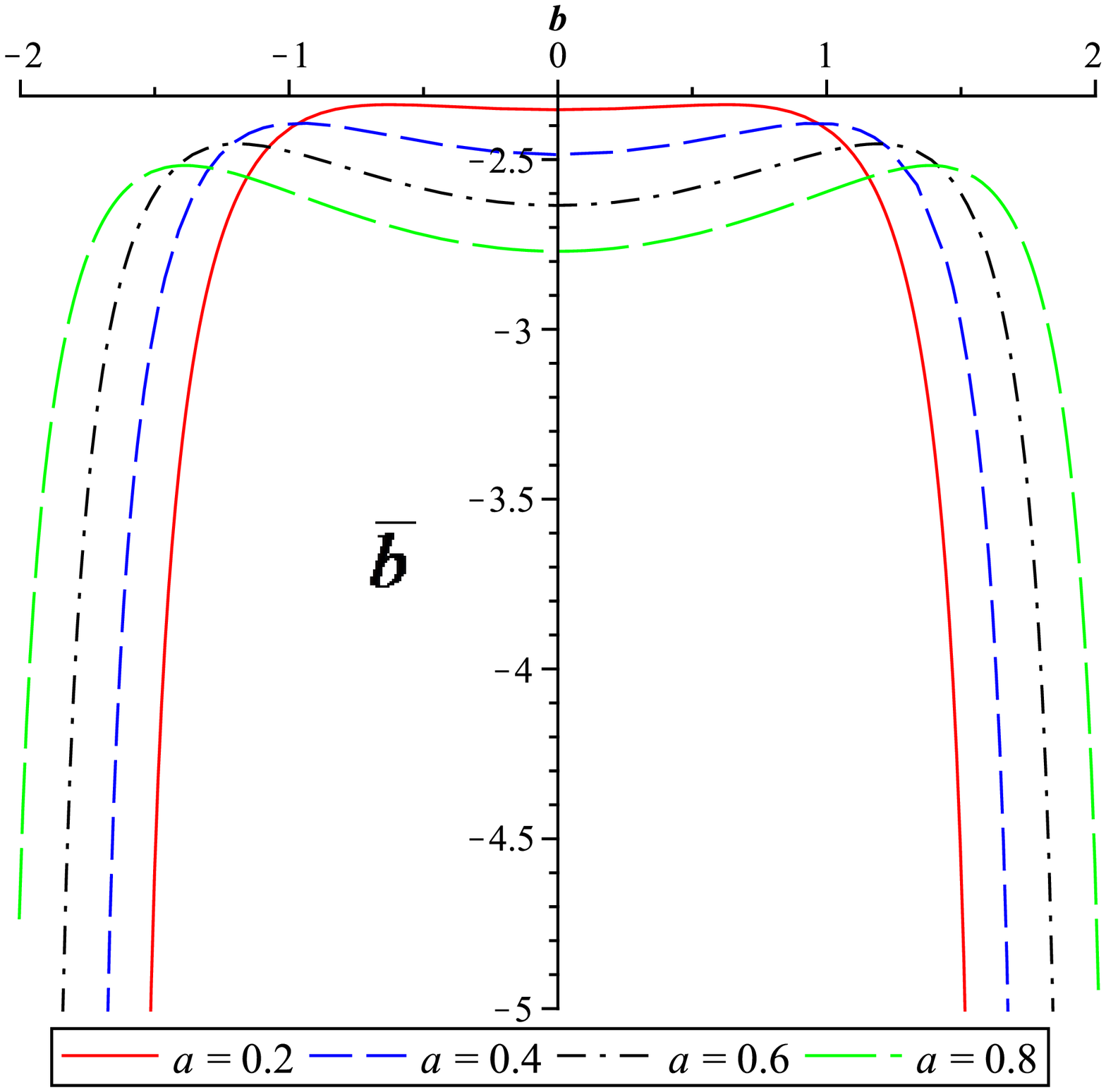}}
\caption{\label{f11} Strong deflection limit coefficient $\bar{a}$, $\bar{b}$ as a function of the spin parameter $b$ for different values of spin parameter $a$ with $m = 1$.}
\end{figure}
We can find the behaviour of the deflection angle very close to the photon sphere following Bozza \cite{vb}. The divergent integral is first split into two parts, one of which contains the divergence and the other is regular. Both these parts are expanded around the radius of photon sphere and approximated with the leading term. We first define two new variables $y$ and $z$ as,
\begin{eqnarray}
 y&=&A(x),\\
 z&=&\frac{y-y_{0}}{1-y_{0}},
\end{eqnarray}
where $y_{0}=A(x_0)$. The total azimuthal angle in terms of two new variables can be expressed as,
\begin{eqnarray}
 \phi(r_{0})=\int_{0}^{1}R(z,x_{0})f(z,x_{0})dz, \label{phiintegral}
\end{eqnarray}
with
\begin{eqnarray}
R(z,x_{0})&=&\frac{2(1-y_{0})}{A'(x)}\frac{\sqrt{B(x)|A(x_0)|}(D(x) + 2 L A(x))}{\sqrt{4 A(x) C^{2}(x) + C(x) D^{2}(x)}},
    \label{Rzr}\nonumber\\
    \end{eqnarray}
 \begin{eqnarray}  
 f(z,x_{0})=\frac{\sqrt{C(x)}}{\sqrt{X}}.\label{fzr}
\end{eqnarray}
Where, $X = C(x) A(x_0) - A(x) C(x_0) + L (A(x) D(x_0)-A(x_0) D(x))$. The function $R(z,x_{0})$ is regular for all the values of $z$ and $x_{0}$, whereas $f(z,x_{0})$ diverges at $z=0$. So, the integral (\ref{phiintegral}) can be separated into two parts
\begin{eqnarray}
 \phi(x_{0})=\phi_{R}(x_{0})+\phi_{D}(x_{0}),
\end{eqnarray}
with
\begin{eqnarray}
 \phi_{D}(x_{0})=\int_{0}^{1}R(0,x_{\textrm{c}})f_{0}(z,x_{0})dz,
\end{eqnarray}
and the regular part
\begin{eqnarray}
 \phi_{R}(x_{0})=\int_{0}^{1}g(z,x_{0})dz,
 \label{regular}
\end{eqnarray}
with $g(z,x_{0}) = R(z,x_{0})\hspace{1mm}f(z,x_{0}) - R(0,x_{\textrm{c}})\hspace{1mm} f_{0}(z,x_{0})$.
In order to find the divergence of the integrand in Eq.(\ref{regular}), we expand the argument of the square root of $f(z,x_{0})$ to second order in $z$:
\begin{eqnarray}
 f(z,x_0)  \sim f_{0}(z,x_{0})=\frac{1}{\sqrt{\alpha z+\beta z^{2}+\mathcal{O}(z^{3})}},
\end{eqnarray}
where
\begin{eqnarray}
\alpha &=& \frac{(1-A(x_0))}{A'(x_0)C(x_0)} \times  \bigg(A(x_0)C'(x_0)-A'(x_0)C(x_0) \nonumber \\ && + L(A'(x_0)D(x_0)
       -A(x_0)D'(x_0))\bigg), \nonumber \\
\end{eqnarray}
\begin{eqnarray}
 \beta&=&\frac{(1-A(x_0))^{2}}{2C(x_0)^{2}A'^{3}(x_0)} \times \nonumber \\
       &&\bigg(2C(x_0)C'(x_0))A'^{2}(x_0)+(C(x_0)C''(x_0)\nonumber\\
       &&-2C'^{2}(x_0))A(x_0)A'(x_0)-C(x_0)C'(x_0)A(x_0)A''(x_0)
            \nonumber\\ 
            && +L\big[A(x_0)C(x_0)(A''(x_0)D'(x_0)-A'(x_0)D''(x_0)) 
             \nonumber\\ &&  +2A'(x_0)C'(x_0)(A(x_0)D'(x_0)-A'(x_0)D(x_0))\big]\bigg).\nonumber\\
\end{eqnarray}
Here, prime denotes the derivative with respect to  $x$. At $x_0=x_{ps}$, $\alpha$ vanishes. The outermost solution of $\alpha=0$ defines the photon sphere.
In the strong deflection limit, the expression for the deflection angle \cite{vb} around the radius of the photon sphere reads as follows,
\begin{eqnarray}
 \alpha(u)=-\bar{a}\log\big(\frac{u}{u(x_p)}-1\big)
                +\bar{b}+\mathcal{O}(u-u(x_p)).\label{Atheta}
\end{eqnarray}
The strong field coefficients $u(x_p)$, $\bar{a}$ and $\bar{b}$ are then given by
\begin{eqnarray}
 u(x_p)&=&L|_{x_{0}=x_{\textrm{p}}},
 \end{eqnarray}
 \begin{eqnarray}
\bar{a}&=&\frac{R(0,x_{\textrm{p}})}{2\sqrt{\beta_p}} 
       =\sqrt{\frac{2A(x_p)B(x_p)}
                   {A(x_p)C(x_p)''-A(x_p)''C(x_p)+
                   u(x_p) \hspace{1mm}Q}},\nonumber\\
\end{eqnarray}
 \begin{eqnarray}
 \bar{b}&=&-\pi+b_{R}+
       \bar{a}\log\bigg(\frac{4\hspace{1mm} \beta_p \hspace{1mm} C(x_p)}
             {u(x_p)|A(x_p)|(D(x_p)
                      +2\hspace{1mm}u(x_p)\hspace{1mm}A(x_p))}\bigg), \nonumber\\
\end{eqnarray}
where, $Q = A(x_p)'' D(x_p) - A(x_p) D(x_p)''$ and $b_R$ goes to zero for the case of 5D MPBH spacetime which is due to the presence of two spin parameters. Here, we have plotted the deflection angle and the parameters $\bar{a}$ and $\bar{b}$, with respect to both the spin parameters $a$ and $b$. In Fig. (\ref{f12}), the deflection angle with respect to spin parameter $b$ for different values of  spin parameter $a$ is graphically presented and it can be observed that the deflection angle first increases followed by a sharp decrease. The decrease in the bending angle depicts the weakening of the force of gravity in the case of 5D MPBH spacetime. It is also observed that in case of Kerr BH, the deflection angle monotonically increases and the variation of deflection angle is different for direct and retrograde photons \cite{iye09}, whereas in case of 5D MPBH, the nature of deflection angle is identical for both, i.e., direct and retrograde photons.\\
The coefficients of strong deflection limit $\bar{a}$ and  $\bar{b}$ are also illustrated in Fig. (\ref{f11}) and one can easily notice that for a fixed value of rotation parameter $a$, $\bar{a}$ increases and $\bar{b}$ decreases with an increase in the value of rotation parameter $b$. One cn also notice that that both the deflection coefficients diverge at the critical point which corresponds to an extremal black hole. We believe that this study might be useful for the investigation of relativistic images in context of mentioned BH spacetime in higher dimension.
\section{Summary and Conclusions}
In this article, we have investigated the frequency shift, cone of avoidance and strong gravitational lensing in the background of a 5D MPBH spacetime. Some of the important results obtained are summarised below.
\begin{itemize}
\item[(i)] The effective potential has a maximum which corresponds to an unstable circular orbit. It is observed that with the increase in the value of spin parameters $a$ and $b$ circular orbit shifts towards and away from the central object respectively.
\item[(ii)] The frequency shift depends on the spin parameters $a$ and $b$. The redshift becomes stronger with the increase in the value of spin parameter $a$ whereas blueshift becomes stronger with the increase in the value of spin parameter $b$. For positive frequency shift the spin parameter $a$ strengthens the gravitational field of a 5D MPBH spacetime.
\item[(iii)] The behaviour of radius of photon sphere indicates that the photons are not easily captured with increasing the value of spin parameter $a$ in case of both type of photon i.e., the direct and retrograde photon.
\item[(iv)] The deflection angle first increases and then decreases with the increase in the value of spin parameter $b$ for fixed values of parameter $a$. There is a significant effect of spin on deflection angle. A decrease in the bending angle shows the decrease in the gravitational strength of the MPBH spacetime. However, the deflection angle and strong field coefficients both changes in a similar way qualitatively with an increase in the spin parameters $a$ and $b$.
\item[(v)] The behaviour of deflection angle and strong field coefficients in 5D MPBH spacetime differs from the Kerr BH spacetime in 4D in GR due to the presence of two spin parameters and the observable quantities are more complex because the spin ($a$ and $b$) breaks the spherical symmetry of the system.\\

We believe that the results obtained herewith would be useful in the study of gravitational lensing
around the rotating BHs in higher dimensions in future.
\end{itemize}
\section*{Acknowledgments}
The authors (RSK and HN) would like to thank the Department of Science and Technology (DST), New Delhi for the financial support through grant no. SR/FTP/PS-31/2009. The author UP would like to thank IUCAA, Pune for academic visits and the Department of Physics, Gurukula Kangri Vishwavidyalay, Haridwar for providing the necessary support during the course of this work been done. The authors (HN and RU) are also thankful to IUCAA, Pune for support under visiting associateship program during their stay at IUCCA, where a part of this study was performed. The authors (RU and HN) would also like to thank the Science and Engineering Research Board (SERB), DST, New Delhi for financial support through the grant number EMR/2017/000339.

\end{document}